\documentclass[12pt,twoside]{article}
\usepackage{amsmath}

\newcommand{\rd}[1]{\mathop{\mathrm{d}#1}}

\newcommand{\fract}[2]{{\textstyle\frac{#1}{#2}}}
\newcommand{\grad}{\vec\nabla}

\newcommand{\CS}{Chern-Simons}
\newcommand{\Cpr}{Clebsch pa\-ra\-me\-ter\-iza\-tion}
\newcommand{\Chg}{Chaplygin gas}
\newcommand{\Gv}{Grass\-mann variables}

\newcommand{\NGa}{Nambu-Goto action}
\newcommand{\pr}{para\-me\-ter\-iza\-tion}

\newcommand{\pp}[1]{\partial_{#1}}
\newcommand{\pa}[1]{\frac\partial{\partial #1}}

\newcommand{\numeq}[2]{\begin{equation}
#2
\label{#1}
\end{equation}}
\newcommand{\refeq}[1]{(\ref{#1})}

\let\vec\boldsymbol

\let\epsilon\varepsilon
\let\phi\varphi

\begin{document}
 
\title{Description of Vorticity by Grassmann Variables\\
and an Extension to Supersymmetry}
\author{R. Jackiw\\
\small\it Center for Theoretical Physics\\ 
\small\it Massachusetts Institute of Technology\\ 
\small\it Cambridge, MA 02139-4307}
\date{\small Typeset in \LaTeX\ by M. Stock\\
MIT-CTP\#3036}
\maketitle

\abstract{\noindent
In this paper particle physics concepts are blended into a field theory for
macroscopic phenomena: Fluid mechanics is enhanced by anticommuting
\Gv\  to describe vorticity, while an additional interaction for the 
\Gv\ leads to supersymmetric fluid mechanics.}

\pagestyle{myheadings}
\markboth{\small {\it R. Jackiw}}{\small  Description of
Vorticity by Grassmann Variables  and an Extension to Supersymmetry}
\thispagestyle{empty}

\section{Pr\'ecis of Fluid Mechanics (With No Vorticity)}

Let me begin with a pr\'ecis of fluid mechanical equations~\cite{ref1}. An
isentropic fluid is described by a matter density field~$\rho$ and a velocity
field~$\vec v$, which satisfy a continuity equation involving the current $\vec j
=\rho\vec v$:
\numeq{eq1}{
\dot\rho + \grad\cdot(\rho\vec v) = 0
}
and a force equation involving the pressure~$P$:
\numeq{eq2}{
\dot{\vec v} + \vec v \cdot\grad\vec v = -\frac1\rho \grad P\ .
}
(Over-dot denotes differentiation with respect to time.) For isentropic fields, the
pressure~$P$ is a function only of the density, and the right side of \refeq{eq2}
may also be written as $-\grad V'(\rho)$, where $V'(\rho)$ is the enthalpy,
$P(\rho) = \rho V'(\rho) - V(\rho)$, and $\sqrt{\rho V''(\rho)} = \sqrt{P'(\rho)}$
is the sound speed (prime denotes differentiation with respect to argument).

Equations~\refeq{eq1} and \refeq{eq2} can be obtained by bracketing the
dynamical variables $\rho$ and $\vec v$ with the Hamiltonian $H(\rho,\vec v)$ 
\numeq{eq3}{
H(\rho,\vec v) = \int \rd r \bigl(\fract12 \rho v^2 + V(\rho)\bigr)
}
\begin{subequations}\label{eq1.4}%
\begin{eqnarray}
\dot\rho &=& \{ H, \rho \} \label{eq4a} \\[1ex]
\dot{\vec v}&=& \{ H, \vec v \} \label{eq4b}
\end{eqnarray}
\end{subequations}%
provided the nonvanishing brackets of the fundamental
variables $(\rho, \vec v)$ are taken to be~\cite{ref2}
\begin{subequations}\label{eq5}%
\begin{eqnarray}
\{ v^i(\vec r), \rho (\vec r') \}&=& \partial_i \delta(\vec r-\vec r')
\label{eq5a} \\[1ex]
\{ v^i(\vec r), v^j (\vec r') \}&=&
-\frac{\omega_{ij}(\vec r)}{\rho(\vec r)} \delta(\vec r-\vec r')\ .
\label{eq5b}
\end{eqnarray}
\end{subequations}%
(The fields in the brackets are at equal times, hence the time
argument is suppressed.)  Here $\omega_{ij}$ is the vorticity, defined as the curl
of $v^i$:
\numeq{eq6}{
\omega_{ij} = \pp i v^j  - \pp j v^i\ .
}

One naturally asks whether there is a canonical 1-form that
leads to the symplectic structure~\refeq{eq5}; that is, one seeks a Lagrangian
whose canonical variables can be used to derive~\refeq{eq5} from canonical
brackets.  When the velocity is irrotational, the vorticity vanishes, $\vec v$ can be
written as the gradient of a velocity potential~$\theta$, $\vec v=\grad\theta$, and
\refeq{eq5} is satisfied by postulating that
\numeq{eq7}{
\{\theta(\vec r), \rho(\vec r') \} = \delta(\vec r-\vec r')
}
that is, the velocity potential is conjugate to the density,
so that the Lagrangian can be taken as
\numeq{eq8}{
L\bigr|_{\mathrm{irrotational}} = \int \rd r\, \theta \dot\rho
 - H\bigr|_{\vec v = \grad\theta}
}
where $H$ is given by \refeq{eq3} with $\vec v=\grad\theta$.

\section{Extending the Formalism to Include Vorticity}

The traditional method of including vorticity in a Lagrangian formalism~\cite{ref3}
involves writing the velocity in a more elaborate potential representation, the
so-called
\Cpr~\cite{ref4}, 
\numeq{eq9}{
\vec v = \grad\theta + \alpha\grad\beta
}
which supports nonvanishing vorticity 
\numeq{eq10}{
\omega_{ij} = \pp i \alpha\, \pp j \beta - \pp j \alpha\, \pp i \beta \ .
}
The Lagrangian
\numeq{eq11}{
L = -\int \rd r \rho(\dot\theta + \alpha\dot\beta) - H
\bigr|_{\vec v = \grad\theta + \alpha\grad\beta}
}
identifies canonical pairs to be $\{\theta,\rho\}$ (as in the irrotational case) and
also $\{\beta,\alpha\rho\}$. It then follows that the algebra \refeq{eq5} is
satisfied, provided $\vec v$ is given by~\refeq{eq9}. The quantities
$(\alpha,\beta)$ are called the ``Gauss potentials''. 

The situation here is similar to the electromagnetic force law: The Lorentz equation
can  be presented in terms of the electric and magnetic field strengths, but a
Lagrangian for the motion requires describing the fields in terms of potentials. 

\section{Some Further Observations on the\\ Clebsch Decomposition of the Vector
 Field $\vec v$}

In three dimensions, \refeq{eq9} involves the same
number of functions on the left and right sides of the
equality: three.  
The total number of dynamical variables $(\rho,\vec v)$ is even -- four -- so an
appropriate phase space can be constructed from the four potentials
$(\rho,\theta,\alpha,\beta)$.
Nevertheless the Gauss potentials are not
uniquely determined by $\vec v$.  The following is the reason
why a canonical formulation of \refeq{eq5} requires using
the Clebsch decomposition  \refeq{eq9}.  Although the
algebra \refeq{eq5} is consistent in that the Jacobi
identity is satisfied, it is degenerate in that the kinematic
helicity $h$
\numeq{eq12}{
h \equiv \fract12 \int \rd{^3 r}  \vec v \cdot (\grad \times
\vec v) =
\fract12 \int \rd{^3 r}   \vec v \cdot \vec \omega
}
($\omega^i=\fract12 \epsilon^{ijk}\omega_{jk}$) has vanishing
bracket with $\rho$ and $\vec v$.  (Note that $h$ is just the
Abelian Chern-Simons term of $\vec v$~\cite{ref5}.)  Consequently, a
canonical formulation requires eliminating the kernel of the
algebra, that is, neutralizing $h$.  This is achieved by the
Clebsch decomposition: $\vec v=\grad \theta+\alpha
\grad\beta$,
$\vec \omega=\grad\alpha\times\grad\beta$, $\vec v
\cdot \vec \omega = \grad\theta \cdot
(\grad\alpha\times\grad\beta) = \grad \cdot
(\theta \grad \alpha \times \grad \beta)$.  Thus in
the Clebsch parameterization the helicity is given by a
surface integral $h = \fract12 \int \rd{\vec S} \cdot
\theta(\grad\alpha\times\grad\beta)$ -- it
possesses no bulk contribution, and the obstruction to a
canonical realization of~\refeq{eq5} is removed~\cite{ref6}.

In two spatial dimensions, the Clebsch parameterization is
redundant, involving three functions to express the two
velocity components.  Moreover, the kernel of \refeq{eq5} 
in two dimensions comprises an infinite number of quantities
\numeq{eq13}{
k_n = \int\rd {^2 r}\rho \Big(\frac{\omega}{\rho} \Big)^n
}
for which the Clebsch parameterization offers no
simplification.  (Here $\omega$ is the two-dimensional
vorticity $\omega_{ij}=\epsilon_{ij}\omega$.) Nevertheless, a
canonical formulation in two dimensions also uses Clebsch
variables to obtain an even-dimensional phase space.

\section{Kinematical Grassmann Variables for Vorticity}

Rather than using the Gauss potentials $(\alpha,\beta)$ of the \Cpr\ \refeq{eq9}
in the description of vorticity \refeq{eq10}, we propose an alternative that makes
use of \Gv~\cite{ref7}. We write
\numeq{eq14}{
\vec v = \grad\theta - \fract12\psi_a \grad \psi_a
}
where $\psi_a$ is a multicomponent, real Grassmann spinor $\psi_a^* = \psi_a$,
$(\psi_a\psi_b)^*= \psi_a^*\psi_b^*$.  (The number of components depends on
spatial dimensionality.) Evidently the nonvanishing vorticity is
\numeq{eq15}{
\omega_{ij} = - \pp i \psi_a\, \pp j \psi_a \ .
}
Moreover, the canonical 1-form in the Lagrangian that replaces \refeq{eq11} reads
\numeq{eq16}{
L = -\int \rd r \rho(\dot\theta -\fract12 \psi_a\dot\psi_a) - H
\bigr|_{\vec v = \grad\theta -\frac12 \psi\grad\psi}
}
The Hamiltonian retains its (bosonic) form \refeq{eq3}, but the \Gv\  are hidden in the formula for the velocity. From the canonical 1-form,
we deduce that $(\theta,\rho)$ remain a conjugate pair [see \refeq{eq7}] and that
the canonically independent \Gv\  are $\sqrt\rho\, \psi$. Thus we
postulate, in addition to the Poisson bracket \refeq{eq7} satisfied by
$(\theta,\rho)$, a Poisson antibracket for the \Gv\ 
\numeq{eq17}{
\{\psi_a(\vec r),  \psi_b(\vec r') \} = -\frac{\delta_{ab}}{\rho(\vec r)} \delta(\vec
r-\vec r') }
and this, together with \refeq{eq7}, has the further consequence that the following
brackets hold:
\begin{gather}
\{\theta(\vec r),  \psi (\vec r') \} = -\frac1{2\rho(\vec r)} \delta(\vec
r-\vec r') \label{eq18}\\
\{\vec v(\vec r),  \psi (\vec r') \} = -\frac{\grad\psi(\vec r)}{\rho(\vec r)}
\delta(\vec r-\vec r')\ . \label{eq19}
\end{gather}
The algebra \refeq{eq5} follows.

One may state that it is natural to describe vorticity by \Gv:
vortex motion is associated with spin, and the Grassmann description of spin
within classical physics is well known. In the model as developed thus far the
Grassmann variables have no role beyond the kinematical one of parameterizing
vorticity~\refeq{eq15} and providing the correct bracket structure. They do
not contribute to the equations of motion for $\rho$ and $\vec v$, \refeq{eq1}
and~\refeq{eq2} [even though they are hidden in the formula~\refeq{eq14} for
$\vec v$]. Moreover, they satisfy a free equation: from~\refeq{eq16}  it  follows
that
\numeq{eq20}{
\dot\psi +\vec v\cdot \grad\psi = 0\ .
}

\section{Dynamical Grassmann Variables for\\ Supersymmetry}

Thus far the \Gv' only role has been to parameterize the velocity/vorticity
\refeq{eq14}, \refeq{eq15} and to provide canonical variables for the
symplectic structure~\refeq{eq5}. The equations for the fluid \refeq{eq1},
\refeq{eq2} are not polluted by them and they do not appear in the Hamiltonian,
beyond their hidden contribution to~$\vec v$. Thus the equation for the
Grassmann fields is free~\refeq{eq20}. 

But now we enquire whether we can add a Grassmann term to the Hamiltonian so
that the \Gv\ enter the dynamics and the entire model enjoys supersymmetry.

We have succeeded for a specific form of the potential $V(\rho)$:
\numeq{eq21}{
V(\rho) = \frac\lambda\rho
}
and for the specific dimensionalities of space-time: (2+1) and (1+1). The reason for
these specificities will be explained in the next Section.

The potential \refeq{eq21}, with $\lambda>0$, leads to negative pressure
\numeq{eq22}{
P(\rho) = \rho V'(\rho) - V(\rho) = -2\lambda/\rho
}
and sound speed
\numeq{eq23}{
s = \sqrt{P'(\rho)} = \sqrt{2\lambda}/\rho
}
(hence $\lambda>0$). This model is called the ``\Chg''. 

 Chaplygin introduced his equation of state as a mathematical approximation to
the physically relevant adiabatic expressions $V(\rho)\propto \rho^n$ with
$n>0$~\cite{ref8}. (Constants are arranged so that the Chaplygin formula is tangent
at one point to the adiabatic profile.)  Also it was realized that certain deformable
solids can be described by the Chaplygin equation of state~\cite{ref9}. These days
negative pressure is recognized as a possible physical effect: exchange forces in
atoms give rise to negative pressure; stripe states in the quantum Hall effect may
be a consequence of negative pressure; the recently discovered cosmological
constant may be exerting negative pressure on the cosmos, thereby accelerating
expansion.

\subsection{Planar model}

In (2+1) dimensions the \Gv\ possess 2-components and two real $2\times2$
Dirac ``$\alpha$''-matrices act on them: $\alpha^1 =\sigma^1$, $\alpha^2=\sigma^3$.
The supersymmetric Hamiltonian is 
\numeq{eq24}{
H = \int \rd{^2r}\Bigl\{ \fract12 \rho v^2 + \frac\lambda\rho +
\frac{\sqrt{2\lambda}}2 \psi\vec\alpha \cdot\grad\psi\Bigr\}
 }
where it is understood that $\vec v = \grad\theta -\frac12
\psi\grad\psi$~\cite{ref7,ref10}. While the continuity equation retains its
form~\refeq{eq1}, the force equation acquires a contribution from the Grassmann
variables
\numeq{eq25}{
\dot{\vec v} + \vec v \cdot \grad \vec v = \grad\frac\lambda{\rho^2} + 
\frac{\sqrt{2\lambda}}\rho (\grad\psi)\vec\alpha \cdot \grad\psi
}
and $\psi$ is no longer free:
\numeq{eq26}{
\dot\psi + \vec v \cdot \grad \psi = \frac{\sqrt{2\lambda}}\rho
 \vec\alpha \cdot \grad\psi\ .
 }
These equations of motion, together with~\refeq{eq1}, ensure that the following
supercharges are time independent:
\begin{subequations}\label{eq27}
\begin{align}
Q &= \int\rd{^2r} \bigl\{ \rho\vec v \cdot(\vec\alpha \psi) +
\sqrt{2\lambda}\psi\bigr\}\label{eq27a}\\
\tilde Q &= \int\rd{^2r} \rho\psi\ .\label{eq27b}
\end{align}
\end{subequations}
 They generate the following transformations:
 \begin{subequations}\label{eq28}
 \begin{alignat}{2}
 \delta\rho &= -\grad\cdot \bigl[ \rho(\eta \vec\alpha\psi)\bigr]  &
 \tilde\delta\rho&=0
 \label{eq28a}\\
 \delta\psi&= -(\eta\vec\alpha\psi) \cdot\grad \psi - \vec
 v\cdot\vec\alpha \eta -\frac{\sqrt{2\lambda}}\rho \eta &\qquad
 \tilde\delta\psi&= -\eta\label{eq28b}\\
 \delta\vec v&= -(\eta\vec\alpha\psi) \cdot\grad v   + \frac{\sqrt{2\lambda}}\rho
 (\eta\grad\psi) &
 \tilde\delta\vec v &= 0\label{eq28c}
 \end{alignat}
 \end{subequations}
where $\eta$ is a two-component constant Grassmann spinor. The antibrackets of
the supercharges produce other conserved quantities:
\begin{subequations}\label{eq29}
 \begin{align}
\{Q_a, Q_b\} &= -2\delta_{ab} H \label{eq29a}\\
\{\tilde Q_a, \tilde Q_b\} &= -\delta_{ab} N \label{eq29b}\\
\{\tilde Q_a, Q_b\} &= \vec\alpha_{ab} \cdot \vec P + \sqrt{2\lambda}\delta_{ab}
\Omega\ . \label{eq29c} 
 \end{align}
\end{subequations}
Here $N$ is the conserved number $\int\rd{^2r} \rho$, $\vec P$ is the conserved
momentum $\int\rd{^2 r} \rho\vec v$,  and $\Omega$ is a center given by the
volume of space $\int\rd{^2r}$. 

\subsection{Lineal model}

In (1+1) dimensions the \Chg\ equation can be written in compact form in terms
of the Riemann coordinates
\numeq{eq30}{
R_{\pm} = v \pm \sqrt{2\lambda}/\rho\ . 
}
Both eqs.~\refeq{eq1} and \refeq{eq2} are equivalent to 
\numeq{eq31}{
\dot R_{\pm} = - R_{\mp} \pa x R_{\pm}\ . 
}
It is known that this system is completely integrable~\cite{ref11}. One hint for this
is the existence of an infinite number of constants of motion: 
\numeq{eq32}{
I_{\pm}^n = \int \rd x \rho(R_{\pm})^n
}
are time-independent by virtue of~\refeq{eq31}.

The supersymmetric Hamiltonian makes use of a real, 1-component Grassmann
field~$\psi$~\cite{ref12}:
\numeq{eq33}{
H = \int \rd x \Bigl( \fract12 \rho v^2 + \frac\lambda\rho +
\frac{\sqrt{2\lambda}}2 \psi \pa x \psi\Bigr)\ .
}
The velocity is given by $v=\pa x \theta -\frac12 \psi\pa x \psi$ and the
equations of motion for the bosonic variables retain the same form as the absence
of~$\psi$, that is, \refeq{eq1}, \refeq{eq2} continue to hold. The Grassmann field
satisfies
\numeq{eq34}{
\dot\psi + R_- \pa x \psi = 0
}
and a general solution follows immediately with the help of~\refeq{eq31}: $\psi$
is an arbitrary function of~$R_+$. 
\numeq{eq35}{
\psi = \Psi(R_+) 
}
Thus the system remains completely integrable.

The supersymmetry charges and transformation laws are obvious dimensional
reductions of \refeq{eq27}--\refeq{eq28}:
\begin{subequations}\label{eq36}
\begin{align}
Q &= \int \rd x \rho R_+ \psi\label{eq36a}\\
\tilde Q &= \int \rd x \rho  \psi\label{eq36b}
\end{align}
\end{subequations}
\begin{subequations}\label{eq37}
\begin{alignat}{2}
\delta\rho &= -\eta \pa x (\rho\psi) & \tilde\delta\rho&=0 \label{eq37a}\\
\delta\psi &= -\eta \psi \psi' - \eta R_+  & \tilde\delta\psi&=-\eta
     \label{eq37b}\\
\delta v &= -\eta (\psi v)' + \eta R_+ \psi' &\qquad \tilde\delta v&= 0\ .
     \label{eq37c}
\end{alignat}
\end{subequations}
The algebra of these is 
\begin{subequations}\label{eq38}
 \begin{align}
\{Q, Q\} &= -2  H \label{eq38a}\\
\{\tilde Q, \tilde Q\} &= -  N \label{eq38b}\\
\{\tilde Q, Q\} &=   P + \sqrt{2\lambda}  \Omega\ . \label{eq38c} 
 \end{align}
\end{subequations}

In view of \refeq{eq35}, we see that evaluating the supercharges $Q$ and $\tilde
Q$ on the solution gives expressions of the same form as the bosonic conserved
charges~\refeq{eq33}.

Indeed, we recognize that two charges in \refeq{eq36} are the first two in an
infinite tower of conserved supercharges, which generalizes the infinite number of
bosonic conserved quantities~\refeq{eq32}:
\numeq{eq39}{
Q_n = \int \rd x \rho R_+^n \psi \ . 
} 

\section{The Origins of Our Models}

We have succeeded in supersymmetrizing a specific model -- the Chaplygin gas --
in specific dimensionalities -- the 2-dimensional plane and the 1-dimensional line
--  leading to nonrelativistic, supersymmetric fluid mechanics in (2+1)- and
(1+1)-dimensional space-time. The reason for these specificities is that both models
descend from Nambu-Goto models for extended systems in a target  space of one
dimension higher than the world volume of the extended object. Specifically, a
membrane in three spatial dimensions and a string in two spatial dimensions,
when gauge-fixed in a light-cone gauge, can be shown to devolve to a bosonic
\Chg\ in two and one spatial dimensions, respectively~\cite{ref13}. The fluid
velocity potential arises from the single dynamical variable in the gauge-fixed
Nambu-Goto theory, namely, the transverse direction variable for the membrane
in space and the string on a plane. Although  purely bosonic \Chg\ models in other
dimensions can devolve from appropriate Nambu-Goto models for extended
objects, for the supersymmetric case we need a superextended object, and these
exist only in specific dimensionalities. In our case it is the light-cone parameterized
supermembrane in (3+1)-dimensional space-time~\cite{ref14} and the superstring
in (2+1)-dimensional space-time~\cite{ref15} that give rise to our planar and lineal
supersymmetric fluid models.

One naturally wonders whether an arbitrary bosonic potential $V(\rho)$ has a
supersymmetric partner in arbitrary dimensions, and this problem is under
further investigation. One promising approach is to consider \pr s of extended
objects other than the light-cone one. It is known that in the purely bosonic case,
other \pr s of the
\NGa s lead to other fluid mechanical models, and this should carry over to a
supersymmetric generalization. 

Incidentally, the existence of Nambu-Goto antecedents of the fluid models that we
have discussed allows one to understand some of their remarkable properties:
complete integrability in the lineal case; existence of further symmetries (which
we have not discussed here) and relation to other models (which devolve from the
same extended system, but are parameterized differently from the light-cone
method)~\cite{ref16}.

\end{document}